\documentclass[letterpaper, 10 pt, conference]{ieeeconf}  

\IEEEoverridecommandlockouts                              

\usepackage{amsmath,amssymb,amsfonts}
\usepackage{algorithmic}
\usepackage{graphicx}
\usepackage{textcomp}
\usepackage{xcolor}
\usepackage{booktabs}
\usepackage{multirow} 
\usepackage{subfig}
\usepackage[justification = centering]{caption}
\usepackage{cite}    
\makeatletter
\let\NAT@parse\undefined
\makeatother
\usepackage{hyperref}
\usepackage{amsmath}
\usepackage{newtxmath}
\def\BibTeX{{\rm B\kern-.05em{\sc i\kern-.025em b}\kern-.08em
    T\kern-.1667em\lower.7ex\hbox{E}\kern-.125emX}}
\title{\LARGE \bf
High Order Collaboration-Oriented Federated Graph Neural Network for Accurate QoS Prediction
}

\author{Zehuan Chen, and Xiangwei Lai, \textit{Member, IEEE}
\thanks{Z. H. Chen, and X. W. Lai are with the College of Computer and Information Science, Southwest University, Chongqing 400715, China (e-mail: chenzehuan00@163.com; laixwcn@gmail.com).
}}
\begin{document}
\begin{onecolumn}
\maketitle
\thispagestyle{empty}
\pagestyle{empty}

\begin{abstract}

Predicting Quality of Service (QoS) data crucial for cloud service selection, where user privacy is a critical concern. Federated Graph Neural Networks (FGNNs) can perform QoS data prediction as well as maintaining user privacy. However, existing FGNN-based QoS predictors commonly implement on-device training on scattered explicit user-service graphs, thereby failing to utilize the implicit user-user interactions. To address this issue, this study proposes a high order collaboration-oriented federated graph neural network (HC-FGNN) to obtain accurate QoS prediction with privacy preservation. Concretely, it magnifies the explicit user-service graphs following the principle of attention mechanism to obtain the high order collaboration, which reflects the implicit user-user interactions. Moreover, it utilizes a lightweight-based message aggregation way to improve the computational efficiency. The extensive experiments on two QoS datasets from real application indicate that the proposed HC-FGNN possesses the advantages of high prediction accurate and privacy protection.
\end{abstract}


\section{INTRODUCTION}

In this era of information overload, each user is constantly surrounded by massive data. With the increasing prevalence of Service-Oriented Architecture (SOA) and the Internet of Services (IoS), the similarity among candidate services has risen significantly. Consequently, selecting an appropriate service to provide the optimal invocation experience has become a challenge in today\textit{'}s landscape. Quality of Service (QoS) is an important non-functional attribute that depends on various factors \cite{xu2025adaptively,yuan2024fuzzy,yuan2023adaptive,chen2024latent,zhou2023cryptocurrency,luo2020position,li2025learning,chen2022differential}. Factors such as user invocation location and network environment adaptability significantly influence performance. They are crucial for the accurate selection of services with similar functionalities \cite{tang2024temporal,xu2025recursion}.

However, with the continuous increase in the number of users and services, the marginalization of computing is becoming increasingly pronounced. It is impractical to call upon all services to monitor the QoS information of each service in real-world application scenarios \cite{chen2024generalized,bi2022two,bi2023two,wu2017highly,zhang2022error,yan2023modified,wu2025data,yang2024latent,zeng2024novel,qin2023parallel}. Numerous existing studies have proposed methods that focus on analyzing historical QoS data and making predictions to provide effective service recommendations to customers. Such approaches have demonstrated significant research efficacy. They rely on the centralized storage of large amounts of client data, which provides substantial coverage for data analysis and model training \cite{wu2022double,xu2023hrst,wu2023robust,li2022diversified,bi2023proximal,shang2021alpha,song2022nonnegative,ding2022highly}. This results in high prediction accuracy and efficiency. In practical application scenarios, the emphasis on privacy protection and data security has become one of the conditions for clients to request services. Under distributed deployment, client possessed limited information data. Specifically, the historical service call records of the client exhibit significant sparsity \cite{b10,bi2023fast,luo2021fast,yuan2020multilayered,wu2024outlier,yuan2022kalman,chen2021hierarchical,chen2024state,qin2023asynchronous,wu2023graph,wu2022multi}. Therefore, accurately predicting the missing information in such a highly sparse user-service matrix is a key issue in QoS estimation \cite{luo2024pseudo,xin2019non,yuan2020generalized,chen2021hyper,li2021proportional,yuan2018effects,li2022momentum,yuan2025proportional,qin2023adaptively,wu2022prediction}.

In previous studies, numerous approaches for distributed processing have been proposed. Federated learning has emerged as a prominent research direction in recent years. As a core technology in machine learning, it addresses the issue of data silos. It ensures data privacy and security while enabling efficient model training. This approach has shown particularly notable performance in several recommendation systems, such as FedGNN \cite{b11} and FedSoG \cite{b12}. However, these methods obtain implicit user-user interaction information either through direct provision by a third-party server or based on shared interaction items. Relying on third-party services to acquire implicit user-user interaction information incurs high costs and poses significant privacy risks. Additionally, treating users with shared interaction items as neighbors cannot guarantee that all implicit user-user interactions will positively impact the prediction performance. Consequently, they cannot be applied to real-world QoS prediction tasks. Recent studies have applied deep learning techniques for QoS prediction \cite{b18,luo2024tsdrl}, effectively learning the complex nonlinear relationships between users and services \cite{li2023nonlinear,liu2023symmetry,chen2025enhancing,zhao2024heterogeneous,li2023nonlinear,CHEN2017128,liu2023high}. However, it relies on local data, making the protection of local information privacy particularly important. To address the aforementioned issues, our paper proposes a high order collaboration-oriented federated graph neural network for accurate QoS prediction. Our contributions can be summarized as follows:
\begin{itemize}
\item We propose a high order collaboration-oriented federated graph neural network (HC-FGNN) for accurate QoS prediction. By integrating lightweight techniques, we optimize the classic federated learning model.
\item In the neighbor aggregation process, we employ explicit user-service graphs, leveraging the principles of the attention mechanism to capture higher-order collaborations. This method reduces the influence of less significant user neighbors on the local model.
\item The experimental results on real datasets indicate that the proposed HC-FGNN model demonstrates a significant improvement in prediction accuracy.
\end{itemize}

\section{RELATED WORKS}

\subsection{Federated Learning}

With the development of internet technologies, an increasing number of users are becoming concerned about privacy. Numerous methods have been proposed for centralized analysis and prediction of historical QoS data, such as QosGNN \cite{b17} and the methods proposed by Ding \textit{et al}. \cite{b18}. However, these approaches do not adequately consider the protection of user privacy, which may lead to users being less willing to share their data, ultimately resulting in decreased model accuracy. The foundational federated learning model (FL) was proposed by Google in 2017. It can effectively protect user privacy, leading to the emergence of an increasing number of federated learning models. For instance, the classic framework FedGNN \cite{b11}. It utilizes graph neural networks to capture the correlations between users and items. Additionally, a federated learning framework for social recommendation FedSoG \cite{b12}. In this approach, users with similar interactions are treated as implicit user-user interaction information. This method may consider users with low similarity as neighbors, potentially affecting the accuracy of the local model. 

\subsection{Graph Neural Network}

In recent years, the development of Graph Neural Networks (GNNs) has been accelerating \cite{yuan2024node,luo2021novel,wang2024gt,b37,bi2023two,yuan2024adaptive,he2024polarized,bi2024graph,lin20243d,chen2022mnl,zhao2025regulation}. Sincebi2024graph Scarselli \textit{et al}. \cite{b21} first introduced the fundamental concept of GNNs, emphasizing their capacity to process graph-structured data, a series of models related to GNNs have been proposed. 
For instance, Kipf and Welling introduced the Graph Convolutional Network (GCN), pioneering convolution-based graph learning methods. The introduction of LightGCN\cite{b14} eliminates the impact of feature transformation and non-linear activation, thereby simplifying the traditional GCN. The advent of Attention Networks (GAT \cite{he2024structure}) and Graph Autoencoders further enhanced the capabilities of graph learning. The emergence of these technologies has significantly advanced the development of GNNs. The application of GNN-related technologies in QoS prediction also presents a promising avenue. Notably, existing studies, including TLGCN \cite{bi2022two} and DeepQSC \cite{b27}, have demonstrated strong performance in QoS prediction tasks.

\section{PROPOSED MODEL}
In this section, we introduce the HC-FGNN model proposed in this paper, providing a detailed explanation of its design principles and overall framework.

\subsection{Problem Formulation} In the context of real QoS data, we primarily consider user IDs, item IDs, and the interaction results that we recognize as indicative of performance. They are represented as users \(U=\{u_1, u_2, u_3, \dots, u_n\}\), items \(I=\{i_1, i_2, i_3, \dots, i_m\}\), and interactions \(R=\{r_{\substack{u1}}, r_{\substack{u2}}, r_{\substack{u3}}, \dots, r_{\substack{us}}\}\), where \textit{n}, \textit{m}, and \textit{s} denote the number of users, items, and interactions, respectively. The corresponding embedding vectors can be represented as \(e_u\in\mathbb{R} ^{n\times d}\), \(e_i\in\mathbb{R} ^{m\times d}\), and \(e_r \in \mathbb{R} ^{n\times m}\), where \textit{d} represents the feature dimension of each embedding vector. In our experiments, this dimension is fixed at 200.

\begin{figure}[!t]\centering
	\includegraphics[width=0.95 \textwidth, bb=0 0 963.6 483.6]{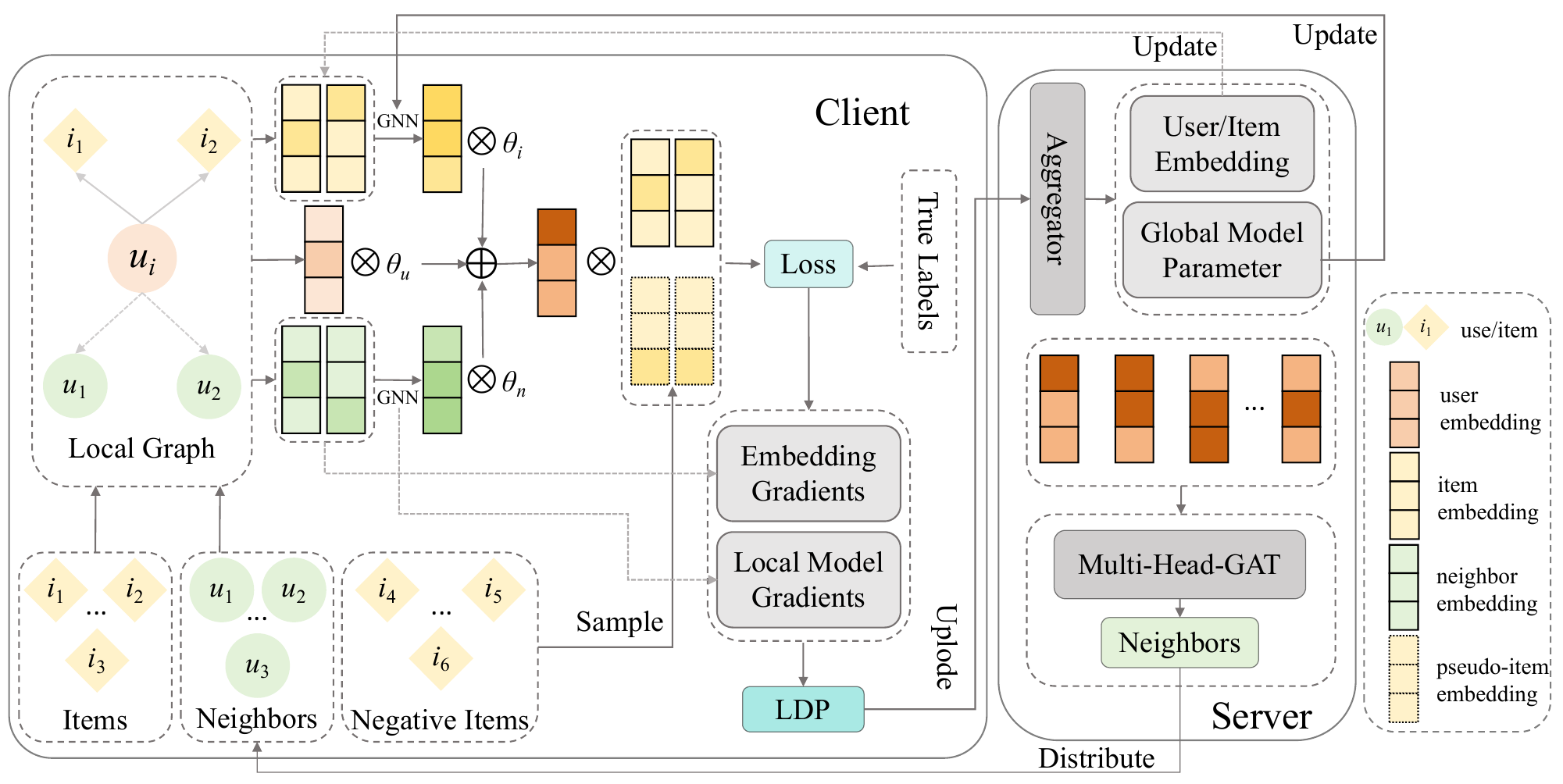}
        \captionsetup{font=footnotesize}
	\caption{Overview of the HC-FGNN. Client: The local model training process will be conducted based on the existing information to update the parameters of the personalized local model. Server: During each training round, the server performs similarity evaluation for each user embedding vector using a multi-head attention mechanism. The multiple similarity scores obtained are averaged to determine the final similarity score, identifying the top-\textit{k} most similar users. These user embeddings are then assigned to the corresponding clients to construct a user relationship graph.}
    \label{FIG_1}
\end{figure} 

\subsection{Overview of the Framework}

This section introduces the proposed framework. As shown in Fig. 1, the overall framework of the model is divided into two parts (the user side and the server side). It is important to note that user data on the client side is private, and data exchange between users is not permitted. The server side holds all the embedding data from the clients and can assess user-user similarity. Ultimately, it selects and assigns an appropriate set of \textit{k} user-user interaction information to each client for constructing the user relationship graph. This approach enriches the information available to each client, enabling personalized local model learning and improving predictive performance.

The specific details are presented in subsections C and D.

\subsection{Higher-Order Neighbor Information Extraction}

Different embeddings \(e_u (e_1, e_2, \dots, e_n)\), for various users exhibit certain similarities. By employing appropriate similarity judgment methods, it is possible to effectively identify the most relevant users. Utilizing this information to construct the user information graph can significantly enhance the high-order collaborative information available for users.

In this paper, we evaluate the similarity of user embeddings transmitted from each client to the server using a multi-head attention mechanism \cite{b28,xu2025attention,li2023saliency}. The specific calculation method for the similarity between user ${i} and user ${j} is illustrated in Equations (1) and (2). 
\begin{equation}
E_{ij}= LeakyReLU ( \alpha \left [ We_{i} \parallel We_{j} ] \right ), \end{equation}
\begin{equation}
\alpha _{ij}= softmax ( E_{ij} ) = \frac{exp\left ( E_{ij} \right ) }{\sum _{k\in N_{i} }exp \left ( E_{ik} \right ) }, 
\end{equation}
where \(e_i\) denotes the embedding of user \textit{i}, \textit{W} represents the learnable weight matrix, \({\alpha}\) is the parameter vector for the attention mechanism, and \textit{exp(·)} indicates the exponential function. \(N_i\) refers to the neighborhood of all nodes adjacent to node \textit{i}. After activation using the \textit{LeakyReLU(·)} function, we obtain the similarity coefficient \(E_{\substack{ij}}\) between feature nodes \textit{i} and \textit{j}. Subsequently, all obtained similarity coefficients undergo \textit{softmax(·)} normalization, resulting in the normalized correlation \(\alpha_{\substack{ij}}\) between the features of nodes \textit{i} and \textit{j}.

The results presented above are from single-head computations. Ultimately, the multi-head computation results will be flattened, and the average value will be taken as the final similarity score. Subsequently, the \textit{k} users with the highest similarity will be selected as neighbor information and transmitted to the corresponding client. The computational flowchart is illustrated in Fig. 2, where Q, K, and V are generated by the linear projection of the feature vectors of the user nodes, \textit{h} represents the number of heads (here example \textit{h}=3), \(n_u\) represents the number of users, \textit{k} represents the number of selected neighbors, and the \textit{d} represents the dimension of the feature vector. In Fig. 2, the left side of the figure illustrates the process flow of the multi-head attention mechanism, while the right side expands on the operations performed on multiple evaluation results.

\subsection{Node Embedding Update}
In existing experimental frameworks, FedSoG performs local aggregation (user, item, and neighbor information) on the client side to update user embedding vectors. This approach effectively considers the impact of multiple relationships on user characteristics, resulting in a rich representation of information at the user end. This enables the upload of valuable information to update the global and local models. Inspired by this model design, we similarly update user features in QoS datasets using local model updates based on user nodes, interaction nodes, and neighbor information. The specific calculation process is as follows:

First, using equations (1) and (2), the attention coefficients between users 
\( 
\left ( u_{i} ,\ u_{j} \right )
\)
and between users and items 
\(
\left ( u_{i} ,\ u_{t} \right )
\)
are calculated, denoted as \(\alpha_{\substack{ij}}\) and \(\beta_{\substack{it}}\), respectively. Multiple neighboring users and items are then aggregated neighbor to\(A_{\substack{un}}\) and item graph embeddings \(A_{\substack{ui}}\), as shown in the following equation (3).
\begin{equation}
A_{un} = \sum_{j= 1}^{n} \alpha _{ij} We_{j}, \ A_{ui} = \sum_{t= 1}^{m} \beta _{it} We_{t}.
\end{equation}

To update the user\textit{'}s own embedding, we utilize equations (1) and (2) to derive the weights \(\theta_n\), \(\theta _i\), and \(\theta_u\), which quantify the importance of the user node concerning itself, its neighbors, and associated items. Accordingly, the embedding representation of the current user node is given by:
\begin{equation}
e_{u}^{'} = \theta _{u} e_{u} + \theta _{i} A_{ui} + \theta _{n} A_{un}.
\end{equation}

Additionally, we can calculate the predicted value by taking the inner product of the user and item embeddings for nodes, as shown in equation (5).
\begin{equation}
\hat{r} _{ui} = e_{u}^{'} \cdot e_{i}.
\end{equation}

To compute the loss, we using the predicted and true values, as shown in (6).
\begin{equation}
L_{u} = \sqrt{\frac{1}{m} \sum_{i= 1}^{m} \left ( r_{ui} - \hat{r} _{ui} \right ) }.
\end{equation}

\subsection{Global Information Aggregation}
In the global aggregation on the server side, an appropriate aggregation strategy should be selected to generate a suitable global model, enabling effective feedback when distributed to clients. In our experiments, when calculating the global loss, we measure the contribution of each user to the global loss based on the number of items they interact with. The global loss is defined equation (7), where \(num_u\) represents the number of items provided by user \textit{u}.
\begin{equation}
L= \sum_{u=1}^{n} num_{u} L_{u}^{2} .
\end{equation}

\subsection{Privacy Protection}
In traditional federated learning, data protection has achieved by clients uploading gradients instead of raw data. However, this approach cannot fully guarantee privacy and security. Because it can still infer the original data from the uploaded information, posing a risk of privacy leakage.

\begin{figure}[t]\centering
	\includegraphics[width=0.8 \textwidth, bb=0 0 865.68 566.14]{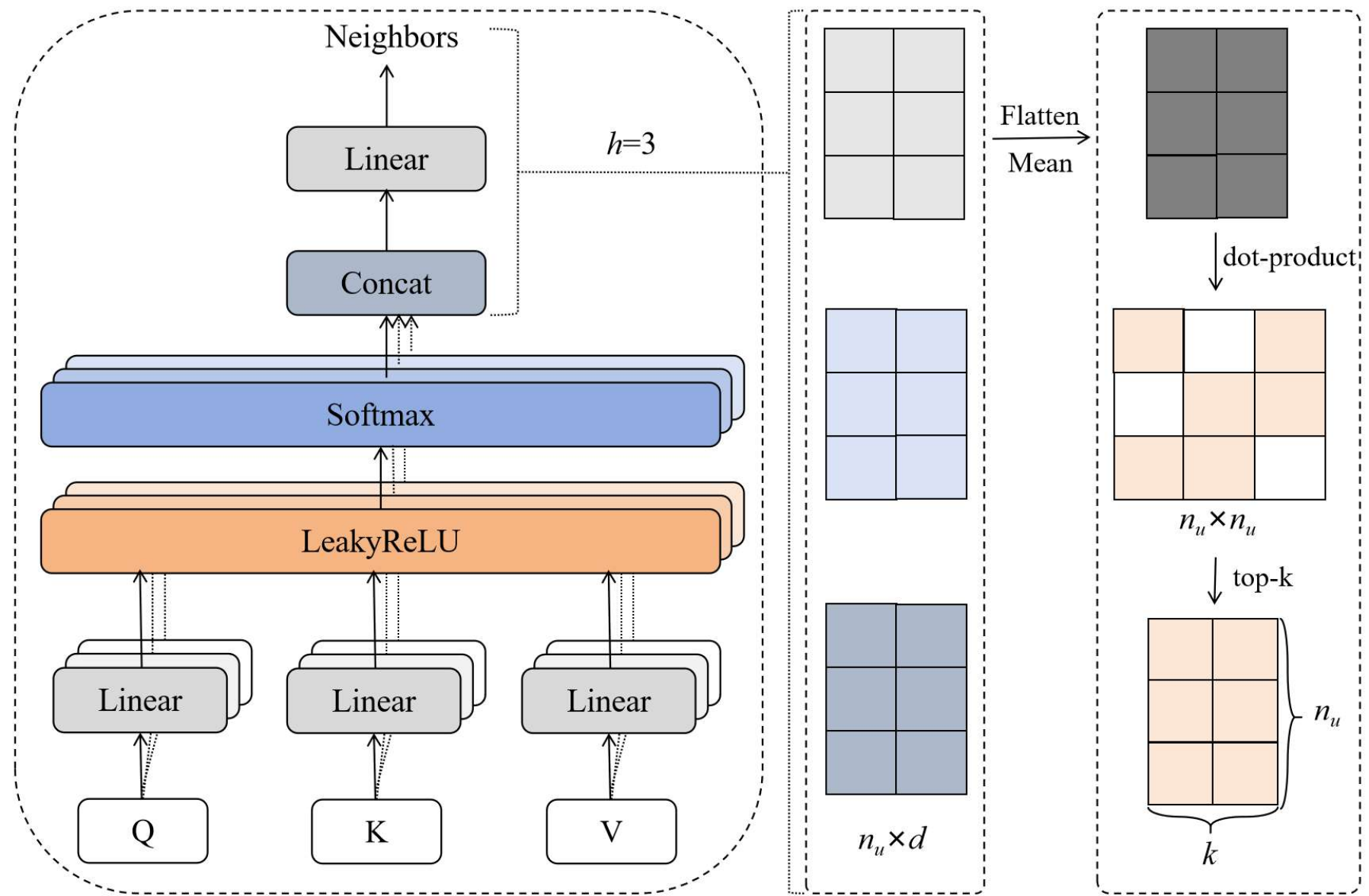}
        \captionsetup{font=footnotesize}
	\caption{Diagram of the Multi-Head Attention Mechanism.}
    \label{FIG_2}
\end{figure}

Our model employs local differential privacy (LDP) techniques to protect user privacy during the gradient upload process. We sample negative samples without interaction terms on the client side and incorporate them into the private data to form new gradient information \textit{g}. By utilizing the information with confounding data as user information and applying LDP to introduce noise, we obtain updated gradient information \textit{g'}. This approach effectively safeguards user privacy while maintaining the model\textit{'}s performance. The specific calculation formulas are as follows:
\begin{equation}
g'= clip\left ( g,\ \delta  \right ) +  laplace\left ( 0,\ \lambda \cdot mean\left ( g \right )  \right ), 
\end{equation}
where \(\delta\) represents the clipping threshold, which restricts the \textit{clip(·)} operation to not exceed the value of \(\delta\). The \(laplace(0, \lambda \cdot mean(g))\) denotes Laplace noise with a mean of 0 and a noise intensity of \(\lambda \cdot mean(g)\). The noise intensity varies with the magnitude of user gradients, thereby adapting to gradients of different scales.

\begin{table}[t]
\centering
\captionsetup{labelsep=newline, font=footnotesize}
\caption{\textsc{Dataset Partitioning}}
\begin{tabular}{@{}cccccc@{}}
\toprule[1.2pt]
\textbf{Datasets} & \textbf{E} & \textbf{U} & \textbf{I} & \textbf{No.} & \textbf{Train:Valid:Test} \\
\midrule[0.8pt]
\multirow{4}{*}{\shortstack{Response \\ Time \\ (RT)}} & \multirow{4}{*}{\parbox{1.3cm}{\centering 1,873,838}} & \multirow{4}{*}{\parbox{0.7cm}{\centering 339}} & \multirow{4}{*}{\parbox{0.7cm}{\centering 5825}}                                                               & RT1 & 1\% : 4\% : 95\% \\
                                            &           &     &      & RT2 & 2\% : 8\% : 90\% \\
                                            &           &     &      & RT3 & 3\% : 12\% : 85\% \\
                                            &           &     &      & RT4 & 4\% : 16\% : 80\% \\ 
\midrule[0.8pt]
\multirow{4}{*}{\shortstack{Throughput \\ (TP)}} & \multirow{4}{*}{\parbox{1.3cm}{\centering 1,831,253}} &   \multirow{4}{*}{\parbox{0.7cm}{\centering 339}}  &   \multirow{4}{*}{\parbox{0.7cm}{\centering 5825}}                                                                         & TP1 & 1\% : 4\% : 95\% \\
                                            &           &     &      & TP2 & 2\% : 8\% : 90\% \\
                                            &           &     &      & TP3 & 3\% : 12\% : 85\% \\
                                            &           &     &      & TP4 & 4\% : 16\% : 80\% \\ 
\bottomrule[1.2pt]
\end{tabular}
\end{table}

\section{EXPERIMENTS}

\subsection{General Settings}

\textbf{Datasets.} We conducted experiments on two real QoS datasets to validate the effectiveness of the proposed model HC-FGNN. Specifically, both datasets are derived from the WSDREAM \cite{bi2022two} system. They are divided into four different ratios, as shown in TABLE I.

\textbf{Evaluation Protocol.} In our experiments, we utilized two evaluation metrics: RMSE and MAE. It is worth noting that smaller values for these metrics correspond to better model performance. Their specific formulas are as follows:
\begin{equation*}
RMSE = \sqrt{\frac{1}{m}\sum_{i=1}^{m} \left ( \hat{r}_{ui} - r_{ui}  \right )^{2} }, \ MAE = \frac{1}{m}\sum_{i=1}^{m}  \left | \left ( \hat{r}_{ui} - r_{ui}  \right )\right |.
\end{equation*}

\textbf{Baselines.} In this paper, we compare some centralized models and some federated models, evaluating the results of different models under the same evaluation metrics. Our objective is to approach the effectiveness of centralized models. The following are the models compared in this study.

\textbf{Centralized Models:} 
\begin{itemize}
\item \textbf{MGDCF}\cite{b33}: MGDCF establishes the equivalence between state-of-the-art GNN-based CF models and a traditional 1-layer NRL model with context encoding.
\item \textbf{PopGo}\cite{b34}: It quantifies the interaction-wise shortcut degrees without any assumptions and mitigates them toward debiased representation learning.
\item \textbf{SGL}\cite{b35}: It supplements the classical supervised task with an auxiliary self-supervised task, which reinforces node representation learning via self-discrimination.
\end{itemize}

\textbf{Federated Learning Models:} 
\begin{itemize}
\item \textbf{FedGNN}\cite{b11}: A privacy-preserving model update method, utilizing a graph expansion protocol with privacy protection facilitated by a third-party server.
\item \textbf{FedSoG}\cite{b12}: A federated learning framework for social recommendation based on graph neural networks. In the initial stage, it provides first-order neighbor information.
\item \textbf{MetaMF}\cite{b36}: A federated meta matrix factorization method for generating personalized item embeddings and recommendation models.
\end{itemize}

\textbf{Hyperparameters.} In HC-FGNN, we set the dimensionality of the latent features to a fixed value of 200. The learning rate \textit{l} is configured to be in the set \{0.01, 0.1\} to accommodate varying densities of the training datasets. The training batch size is set to \{32, 64, 128, 256, 339\} to examine the impact of different batch sizes on training accuracy. The number of neighbors \textit{k} is set to \{5, 10, 20, 40, 80\} to control the quantity of user neighbors. The regularization coefficient \(l_2\) is set to \{0.01, 0.1\} to mitigate overfitting and enhance the model\textit{'}s generalization capability.
\begin{table*}[htbp]
\centering
\captionsetup{labelsep=newline, font=footnotesize}
\caption{\textsc{Performance Comparison Across Methods}}
{ 
\begin{tabular}{llccc|cccc}
\toprule[1.2pt]
\textbf{No.} & \textbf{Metric} & \textbf{MGDCF} & \textbf{PopGo} & \textbf{SGL} & \textbf{FedGNN} & \textbf{FedSoG} & \textbf{MetaMF} & \textbf{HC-FGNN} \\ \midrule[0.8pt] 
\multirow{2}{*}{RT1}  & RMSE                   & \underline{1.889\(_{\pm3\mathrm{E}-3}\)} & 2.069\(_{\pm5\mathrm{E}-3}\) & 1.901\(_{\pm4\mathrm{E}-3}\) & 2.304\(_{\pm7\mathrm{E}-3}\) & 2.171\(_{\pm5\mathrm{E}-3}\) & \textbf{1.967\(_{\pm7\mathrm{E}-3}\)} & 2.003\(_{\pm5\mathrm{E}-3}\) \\
                      & MAE                    & 0.753\(_{\pm2\mathrm{E}-2}\) & 0.942\(_{\pm6\mathrm{E}-3}\) & \underline{0.748\(_{\pm2\mathrm{E}-3}\)} & 1.223\(_{\pm6\mathrm{E}-3}\) & 0.908\(_{\pm6\mathrm{E}-4}\) & 0.945\(_{\pm4\mathrm{E}-4}\) & \textbf{0.817\(_{\pm3\mathrm{E}-3}\)} \\ 
\multirow{2}{*}{RT2}  & RMSE                   & \underline{1.687\(_{\pm4\mathrm{E}-3}\)} & 1.872\(_{\pm2\mathrm{E}-3}\) & 1.714\(_{\pm2\mathrm{E}-3}\) & 2.233\(_{\pm9\mathrm{E}-3}\) & 2.171\(_{\pm8\mathrm{E}-3}\) & 1.978\(_{\pm6\mathrm{E}-3}\) & \textbf{1.968\(_{\pm8\mathrm{E}-3}\)} \\
                      & MAE                    & \underline{0.694\(_{\pm3\mathrm{E}-3}\)} & 0.851\(_{\pm3\mathrm{E}-3}\) & 0.721\(_{\pm4\mathrm{E}-3}\) & 1.119\(_{\pm5\mathrm{E}-3}\) & 0.907\(_{\pm5\mathrm{E}-4}\) & 0.966\(_{\pm5\mathrm{E}-4}\) & \textbf{0.761\(_{\pm4\mathrm{E}-3}\)} \\ 
\multirow{2}{*}{RT3}  & RMSE                   & \underline{1.617\(_{\pm2\mathrm{E}-3}\)} & 1.733\(_{\pm4\mathrm{E}-3}\) & 1.657\(_{\pm3\mathrm{E}-3}\) & 2.224\(_{\pm6\mathrm{E}-3}\) & 2.170\(_{\pm4\mathrm{E}-3}\) & 1.977\(_{\pm5\mathrm{E}-3}\) & \textbf{1.931\(_{\pm3\mathrm{E}-3}\)} \\
                      & MAE                    & \underline{0.692\(_{\pm3\mathrm{E}-3}\)} & 0.795\(_{\pm6\mathrm{E}-3}\) & 0.730\(_{\pm2\mathrm{E}-3}\) & 1.107\(_{\pm4\mathrm{E}-3}\) & 0.907\(_{\pm1\mathrm{E}-4}\) & 0.977\(_{\pm2\mathrm{E}-3}\) & \textbf{0.759\(_{\pm5\mathrm{E}-3}\)} \\ 
\multirow{2}{*}{RT4}  & RMSE                   & \underline{1.554\(_{\pm2\mathrm{E}-3}\)} & 1.657\(_{\pm5\mathrm{E}-3}\) & 1.656\(_{\pm3\mathrm{E}-3}\) & 2.226\(_{\pm8\mathrm{E}-3}\) & 2.173\(_{\pm7\mathrm{E}-3}\) & 1.975\(_{\pm6\mathrm{E}-3}\) & \textbf{1.906\(_{\pm6\mathrm{E}-3}\)} \\
                      & MAE                    & \underline{0.690\(_{\pm4\mathrm{E}-3}\)} & 0.778\(_{\pm4\mathrm{E}-3}\) & 0.734\(_{\pm5\mathrm{E}-3}\) & 1.107\(_{\pm4\mathrm{E}-3}\) & 0.908\(_{\pm3\mathrm{E}-4}\) & 1.020\(_{\pm5\mathrm{E}-3}\) & \textbf{0.756\(_{\pm3\mathrm{E}-3}\)} \\ 
\multirow{2}{*}{TP1}  & RMSE                   & \underline{0.916\(_{\pm5\mathrm{E}-4}\)} & 1.110\(_{\pm9\mathrm{E}-3}\) & 0.977\(_{\pm1\mathrm{E}-2}\) & 1.286\(_{\pm2\mathrm{E}-3}\) & 1.206\(_{\pm3\mathrm{E}-4}\) & 1.107\(_{\pm2\mathrm{E}-3}\) & \textbf{1.062\(_{\pm2\mathrm{E}-3}\)} \\
                      & MAE                    & \underline{0.344\(_{\pm4\mathrm{E}-3}\)} & 0.479\(_{\pm3\mathrm{E}-3}\) & 0.400\(_{\pm4\mathrm{E}-3}\) & 0.691\(_{\pm6\mathrm{E}-3}\) & 0.475\(_{\pm2\mathrm{E}-4}\) & 0.543\(_{\pm3\mathrm{E}-3}\) & \textbf{0.414\(_{\pm4\mathrm{E}-3}\)} \\ 
\multirow{2}{*}{TP2}  & RMSE                   & \underline{0.776\(_{\pm5\mathrm{E}-4}\)} & 0.871\(_{\pm7\mathrm{E}-3}\) & 0.925\(_{\pm3\mathrm{E}-3}\) & 1.282\(_{\pm7\mathrm{E}-3}\) & 1.206\(_{\pm3\mathrm{E}-4}\) & 1.108\(_{\pm4\mathrm{E}-3}\) & \textbf{0.947\(_{\pm3\mathrm{E}-3}\)} \\
                      & MAE                    & \underline{0.313\(_{\pm1\mathrm{E}-3}\)} & 0.414\(_{\pm4\mathrm{E}-3}\) & 0.412\(_{\pm4\mathrm{E}-3}\) & 0.695\(_{\pm4\mathrm{E}-3}\) & 0.476\(_{\pm4\mathrm{E}-4}\) & 0.522\(_{\pm9\mathrm{E}-4}\) & \textbf{0.361\(_{\pm2\mathrm{E}-3}\)} \\ 
\multirow{2}{*}{TP3}  & RMSE                   & \underline{0.729\(_{\pm5\mathrm{E}-4}\)} & 1.206\(_{\pm9\mathrm{E}-3}\) & 0.912\(_{\pm5\mathrm{E}-3}\) & 1.287\(_{\pm6\mathrm{E}-3}\) & 1.206\(_{\pm7\mathrm{E}-3}\) & 1.109\(_{\pm4\mathrm{E}-3}\) & \textbf{0.936\(_{\pm1\mathrm{E}-3}\)} \\
                      & MAE                    & \underline{0.285\(_{\pm7\mathrm{E}-4}\)} & 0.475\(_{\pm6\mathrm{E}-4}\) & 0.419\(_{\pm4\mathrm{E}-3}\) & 0.699\(_{\pm9\mathrm{E}-3}\) & 0.476\(_{\pm5\mathrm{E}-4}\) & 0.527\(_{\pm3\mathrm{E}-4}\) & \textbf{0.348\(_{\pm3\mathrm{E}-3}\)} \\ 
\multirow{2}{*}{TP4}  & RMSE                   & \underline{0.471\(_{\pm2\mathrm{E}-4}\)} & 1.224\(_{\pm4\mathrm{E}-3}\) & 0.481\(_{\pm1\mathrm{E}-9}\) & 1.218\(_{\pm6\mathrm{E}-3}\) & 1.224\(_{\pm3\mathrm{E}-4}\) & 1.107\(_{\pm3\mathrm{E}-4}\) & \textbf{0.919\(_{\pm3\mathrm{E}-3}\)} \\
                      & MAE                    & \underline{0.236\(_{\pm5\mathrm{E}-3}\)} & 0.481\(_{\pm6\mathrm{E}-3}\) & 0.402\(_{\pm7\mathrm{E}-4}\) & 0.537\(_{\pm5\mathrm{E}-4}\) & 0.480\(_{\pm5\mathrm{E}-4}\) & 0.535\(_{\pm9\mathrm{E}-4}\) & \textbf{0.345\(_{\pm5\mathrm{E}-3}\)} \\ 
\bottomrule[1.2pt]
\end{tabular}}
\end{table*}

\subsection{Comparison and Analysis of Experimental Results}

The two real-world datasets used in this paper are applied to all comparison models, and the resulting test outcomes are shown in TABLE II. In TABLE II, underlined values indicate the best scores within the centralized models. Bold values represent the best scores within the federated models. From it, we can conclude that:

\begin{enumerate}
\item The proposed model framework achieves favorable experimental results in most federated learning models.
\item The proposed model avoids the cost increase associated with third-party server involvement in FedGNN and simultaneously reduces the risk of privacy leakage.
\item Compared to model that treat all users with the same interactions as neighbors, the proposed model is able to better leverage data with higher similarity. To prevent low-similarity users from impacting the final accuracy.
\item Compared to centralized models, our model achieves a better trade-off between privacy protection and performance, even surpassing some centralized models.
\end{enumerate}

\subsection{Sensitivity Analysis}\label{AAA}
For all hyperparameters, the choice of their values can significantly impact the model\textit{'}s prediction performance. In this study, we focus on discussing the impact of the number of implicit user-user interactions \textit{k} and the training batch size on the prediction performance. We conduct experiments on the RT1 dataset. In our HC-FGNN method, we analyze and present the impact of different values of the two hyperparameters on model performance.

As shown in the Fig. 3., the optimal value of
\textit{k} for the selected dataset (RT1) is approximately \textit{k}=5. As the \textit{k} value increases, the model\textit{'}s performance exhibits fluctuations and degrades significantly when \textit{k} exceeds 20. This indicates that the choice of \textit{k} has a substantial impact on the model\textit{'}s performance. Selecting an excessively large number of implicit user-user interactions results in the inclusion of many low-similarity users as neighbors, which are transmitted to the client and adversely affect the model\textit{'}s predictive performance.

As shown in Fig. 4, the model\textit{'}s performance initially improves with an increasing batch size, reaching its peak at a batch size of 128. However, performance declines when the batch size exceeds 128. Therefore, the optimal batch size for model training in the RT1 dataset is 128.

\subsection{Ablation Study}
In this paper, we magnifies the explicit user-service graphs following the principle of attention mechanism to obtain the high order collaborative information, which reflects the implicit user-user interactions to enrich node information. To validate the effectiveness of the method presented in the proposed framework HC-FGNN, we conduct experiments under identical settings. 

As shown in TABLE III, compared to approaches that do not utilize attention mechanism to construct user information graphs (HC-FGNN (w/o)), the proposed method achieves superior performance. In the RT1 dataset, RMSE and MAE are reduced by 7.23\% and 10.07\%, respectively. In the TP1 dataset, RMSE and MAE are reduced by 11.07\% and 12.52\%, respectively. As observed, the results highlights the importance of effectively constructing high order collaborative information in QoS prediction models. Compared to methods that treat all users with the same interaction items as neighbors (HC-FGNN (all)), our approach achieves notable improvements. On the RT1 dataset, the proposed method reduces RMSE and MAE by 5.48\% and 6.69\%, respectively. Similarly, on the TP1 dataset, RMSE and MAE are reduced by 7.23\% and 6.05\%, respectively. 

\begin{figure}[!t]
\centering
\includegraphics[width=0.7 \textwidth, bb=0 0 685.714 268.857]{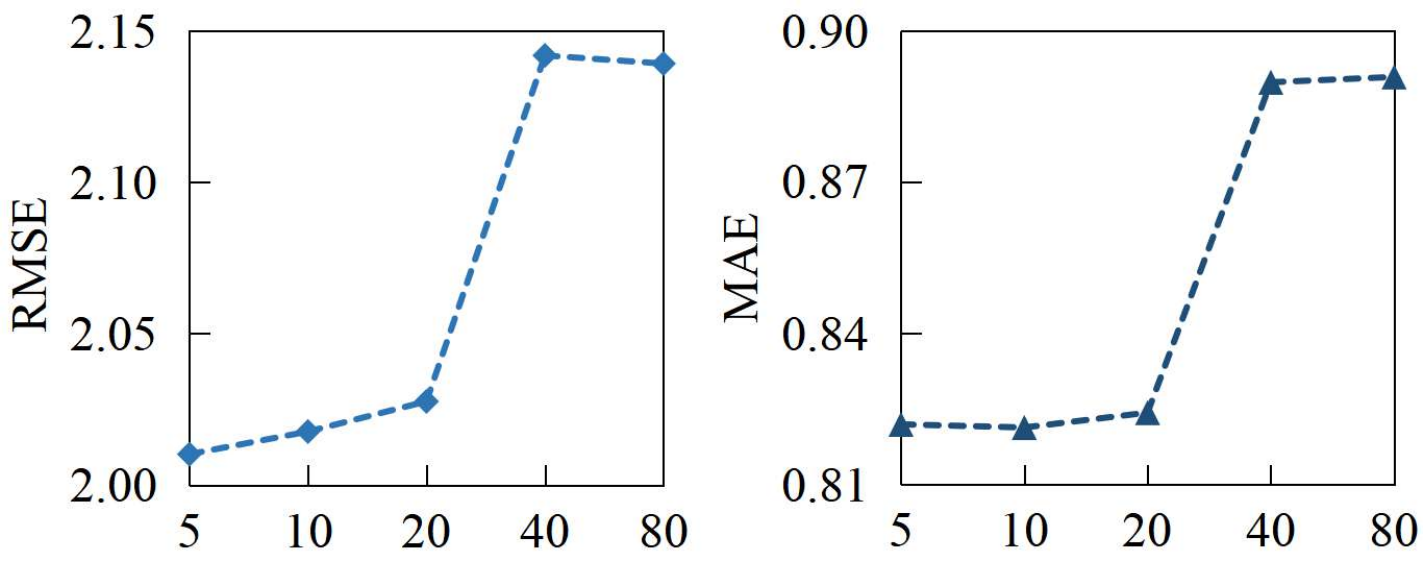}
\captionsetup{font=footnotesize}
\caption{Effect of Varying \textit{k} in the RT1 Dataset.}
\end{figure}

\begin{figure}[!t]
\centering
\includegraphics[width=0.7 \textwidth, bb=0 0 507.29 195.75]{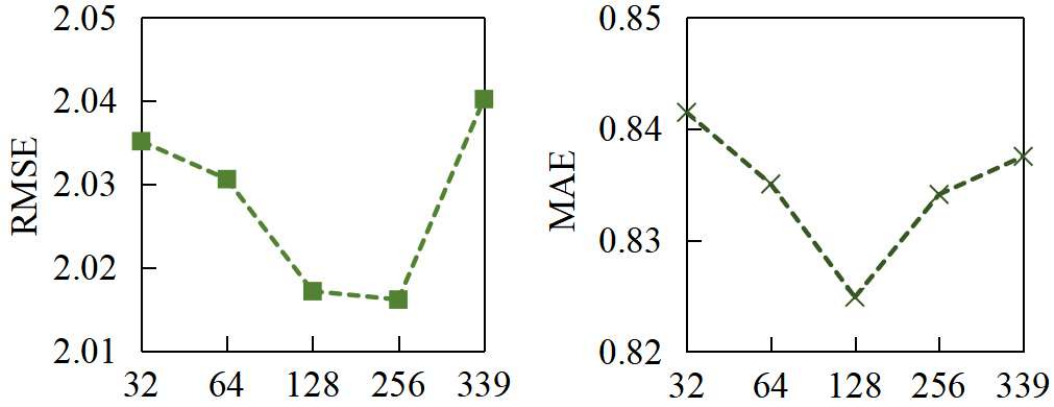}
\captionsetup{font=footnotesize}
\caption{Effect of Varying batch in the RT1 Dataset.}
\end{figure}

\begin{table}[t!]
\centering
\captionsetup{labelsep=newline, font=footnotesize}
\caption{\textsc{Comparison w or w/o higher-order information}}
\begin{tabular}{@{}ccccc@{}}
\toprule[1.2pt]
\multirow{2}{*}{\textbf{Methods}} & \multicolumn{2}{c}{\textbf{RT1}} & \multicolumn{2}{c}{\textbf{TP1}} \\ 
\cmidrule(lr){2-3} \cmidrule(lr){4-5}
                                   & \textbf{RMSE}     & \textbf{MAE}     & \textbf{RMSE}     & \textbf{MAE}     \\ 
\midrule[0.8pt]
HC-FGNN (w/o)              & 2.1588            & 0.9084           & 1.1947            & 0.4729           \\
HC-FGNN (all)               & 2.1190            & 0.8755           & 1.1460            & 0.4404           \\
\textbf{HC-FGNN}                     & \textbf{2.0028}   & \textbf{0.8169}  & \textbf{1.0621}   & \textbf{0.4137}  \\ 
\midrule[0.8pt]
\textit{Improve (vs. w/o)}         & 7.23\%            & 10.07\%          & 11.07\%           & 12.52\%          \\
\textit{Improve (vs. all)}         & 5.48\%            & 6.69\%           & 7.32\%            & 6.05\%           \\ 
\bottomrule[1.2pt]
\end{tabular}
\end{table}

\section{CONCLUSIONS}

We present a high order collaboration-oriented federated graph neural network for accurate QoS prediction, referred to as HC-FGNN. This method enables similarity assessment on the server side based on user-uploaded feature information, allowing for the selection of the top-\textit{k} most similar users as neighbor information. This approach provides higher-order user information at the client side, rather than relying solely on individual client information. Our approach effectively preserves user data privacy while ensuring prediction accuracy. Extensive experiments demonstrate that our proposed HC-FGNN framework achieves commendable performance, and we hope to apply this model to other collaborative recommendation systems, yielding positive outcomes.
\bibliographystyle{ieeetr}
\bibliography{example}
    
\end{onecolumn}
\end{document}